\documentclass[12pt]{article}

\usepackage{amsmath}

\newcommand{\be}{\begin{equation}}
\newcommand{\ee}{\end{equation}}
\newcommand{\ben}{\begin{eqnarray}}
\newcommand{\een}{\end{eqnarray}}
\newcommand{\ba}{\begin{eqnarray}}
\newcommand{\ea}{\end{eqnarray}}
\newcommand{\nn}{\nonumber \\}
\newcommand{\bi}{\begin{itemize}}
\newcommand{\ei}{\end{itemize}}

\newcommand{\lb}{\left (}
\newcommand{\rb}{\right )}
\newcommand{\ltb}{\left [}
\newcommand{\rtb}{\right ]}

\begin{document}
\begin{center}

\vspace{24pt} { \large \bf Schwarzschild-Tangherlini quasinormal modes at large $D$ revisited} \\

\vspace{30pt}

\vspace{30pt}

\vspace{30pt}

{\bf Amruta Sadhu\footnote{sadhuamruta@students.iiserpune.ac.in}}, {\bf Vardarajan
Suneeta\footnote{suneeta@iiserpune.ac.in}}

\vspace{24pt} 
{\em  The Indian Institute of Science Education and Research (IISER),\\
Pune, India - 411008.}

\end{center}
\date{\today}
\bigskip

\begin{center}
{\bf Abstract} 
\end{center}
The large dimension ($D$) limit of general relativity has been used
in problems involving black holes as an analytical approximation tool.
Further it has been proposed that both linear and nonlinear problems
involving black holes can be systematically studied in a $1/D$ expansion.
Certain quasinormal modes of higher-dimensional
Schwarzschild black holes with $\omega \sim \mathcal{O}(1)$ were studied in the large $D$ limit
using a $1/D$ expansion for the mode function.

In this paper, we revisit this linear perturbation problem and obtain
an analytical equation for the vector quasinormal modes $\omega \sim \mathcal{O}(1)$ in the large $D$ limit, without using a $1/D$ expansion
for the mode function. This can be used to compute quasinormal modes to next to leading order in $1/D$. 
We also compute vector and scalar quasinormal modes
with $\omega \sim \mathcal{O}(D)$ in the Laplace transform approach 
in the large $D$ limit. We discuss the useful features of this approach specifically in the
large $D$ limit. 

\newpage
\section{Introduction}

Recent studies of black holes/branes have employed the large dimension $D$ limit to address classical stability issues \cite{kol}---\cite{chen}.
Specifically, the large $D$ limit has been used to study the equations for linearized perturbations of black holes, analyze stability and to compute quasinormal modes. Motivated by this, an effective membrane approach has also been initiated in \cite{emparanrest}, \cite{minwalla} to study large $D$ black holes systematically in a $1/D$ expansion. Possible applications include the nonlinear regime in problems involving black holes (such as mergers) by using the large $D$ limit. 

While the nonlinear regime is difficult to study even in the large $D$ limit, one can certainly address problems in linearized perturbation theory for 
large $D$ Schwarzschild-Tangherlini black holes. In particular, quasinormal modes, which characterize the response of a final black hole after a nonlinear process
such as a merger, can be studied in the large $D$ limit. The hope is that this limit would allow an analytical computation of these modes, and 
a systematic procedure of incorporating $1/D$ corrections.
Emparan and collaborators \cite{emparan} in a $1/D$ expansion for the quasinormal mode functions and the quasinormal mode frequencies, computed a set of modes with frequency $\omega \sim \mathcal{O}(1)$.
In fact, two distinct sets of modes were identified in the large $D$ limit - 
one set with frequency $\omega \sim \mathcal{O}(D)$ and the other set with frequency $\omega \sim \mathcal{O}(1)$, the mode functions for the latter decaying 
away from the horizon.

In our paper, we evaluate the vector quasinormal modes of Schwarzschild-Tangherlini black holes in the large $D$ limit --- in particular, without a $1/D$ expansion
for the quasinormal mode functions for modes with frequency $\omega \sim \mathcal{O}(1)$. The aim is to understand the nature of the 
large $D$ limit of the perturbation equation and its solutions and to ascertain if the solutions admit a $1/D$ expansion. 
In section II, we first find the leading order 
result for vector quasinormal modes with
frequency $\omega \sim \mathcal{O}(D)$ using a Laplace transform approach. We discuss its specific advantages for computing quasinormal modes
in the large $D$ limit. We also  discuss the obstructions to going beyond the leading order in $D$ in our method. We then obtain an equation obeyed by
modes with frequency $\omega \sim \mathcal{O}(1)$. To leading order and the next, our answer for the quasinormal modes agrees with numerical \cite{dias} and analytical
results \cite{emparan}.  
We find that the perturbation equation for $\omega \sim \mathcal{O}(1)$ is a hypergeometric equation which
reduces to a degenerate case in the $D \to \infty$ limit and its general solution cannot be obtained from 
solutions to the degenerate 
case in a $1/D$ expansion --- however, the solution ingoing at the horizon can. 
 In section III and Appendix C, we discuss the perturbation equation for scalar quasinormal modes. 
For frequencies $\omega \sim \mathcal{O}(1)$, this is a Heun equation in which two of the singular points merge in the $D \to \infty$ limit. Due to this,
a direct computation of the quasinormal modes without a $1/D$ expansion is not possible. 
We discuss the leading order scalar quasinormal modes with $\omega \sim \mathcal{O}(D)$.

A summary of results and a discussion of these points is presented in section IV.
Appendices A and B contain some of the computational details of sections II and III.

\section{Vector quasinormal modes}

We now consider perturbations of $D$ dimensional Schwarzschild-Tangherlini black holes. The black hole metric is
\begin{equation}\label{vqnm0}
g_{AB}dx^A dx^B= - f(r) dt^2 + f^{-1} (r)dr^2 +r^2 d\Omega^{2}_{n} ;
\end{equation}
where $f (r) = \left(1-\frac{b^{D-3}}{r^{D-3}}\right)$ and $r \geq b$. 
The event horizon is located at $r=b $. $d\Omega^2_n=\gamma_{ij} d\tilde{y}^i d\tilde{y}^j$ is 
the metric of a $n$-dimensional sphere of unit radius and $n=D-2$.

We consider perturbations of the metric (\ref{vqnm0}), with the perturbed metric $\bar{g}_{AB}=g_{AB}+ h_{AB}$, in linearized perturbation theory. 
We assume the mode decomposition $h_{AB} = \tilde h_{AB} e^{i \omega t}$ where
$\tilde h_{AB}$ is time-independent. The standard procedure consists of classifying perturbations $\tilde h_{AB}$ as tensor, scalar and vector depending on whether they are proportional to the tensor,scalar or vector spherical harmonics (on the $n$-sphere) respectively. Using appropriately defined master variables in each class of perturbations, Ishibashi and Kodama \cite{ishikodama} obtained decoupled equations for each class in a simple Schrodinger form. Denoting the master variable corresponding to vector perturbations as $\Psi_V $, the equation governing vector perturbations is
\begin{equation}
\frac{d^2}{dr_{*}^{2}} \Psi_V + (\omega^2 - V_{V} ) \Psi_V = 0.
\label{vqnm1}
\end{equation}
Here $dr_{*} = \frac{dr}{f(r)}$ and 
\begin{eqnarray}
V_{V} = \frac{(D-3)^2 f(r)}{4 r^2 } \left [ \lb 1 + \frac{2\ell}{D-3} \rb ^2 - \frac{1}{(D-3)^2}
- 3 \lb 1  + \frac{1}{D-3} \rb ^2 \lb \frac{b}{r}\rb ^{D-3} \right ] .
\label{vqnm2}
\end{eqnarray}

For future reference, we note the various distinct regions that appear in this perturbation 
problem in the large $D$ limit. The function $f(r)$ increases steeply from zero 
in $b < r < b + \frac{b}{D}$ 
and is almost constant outside this region. We therefore 
define as the `far' region, $r >> b + \frac{b}{D}$ where the metric is almost flat.             
The potential $V_V $ has its maximum
at $r \sim b + c \frac{\ln D}{D}$ where $c$ is a $D$-independent constant. Also,  
following the authors in \cite{emparan} we can define a near zone by $(r-b) \ll b$. Thus, the
so defined near region and the far region have an overlap for 
$\frac{b}{D} < (r-b) < b$ and the maximum of the potential lies in this region. The
overlap region is significant only for large $D$. We also 
note here that when $b < r << b + \frac{b}{D}$, the $t-r$ part of the black hole metric can be approximately written
in Rindler form. So this is what would be traditionally called  {\em `near-horizon'}. In this paper,
we retain the terminology in \cite{emparan} for our definition of the near region $(r-b) \ll b$ which, for
large $D$ is much bigger than the traditional near-horizon region. We
keep in mind that this definition of the near and far regions with an overlap region is 
a convenient
analytical approximation tool for large $D$.

We change to coordinate $R = (\frac{r}{b})^{D-3}$. 
From (\ref{vqnm1}), the equation for $\Psi_{V} (R)$ is:
\begin{eqnarray}
&&R(1-R) \Psi_{V}''(R) + \left[ -R + \frac{R-1}{D-3} \right] \Psi_{V}'(R) 
+ \left [ \frac{1}{4} \lb 1 + \frac{2\ell}{D-3} \rb ^2 -  \frac{1}{4}\frac{1}{(D-3)^2} \right. \nonumber \\
&& \left. - \frac{3}{4R} \lb 1  + \frac{1}{D-3} \rb ^2 - \frac{\omega^2 r^2 }{(D-3)^2 }
- \frac{\omega^2 r^2 }{(D-3)^2 (R-1)} \right ] \Psi_V~~~~ \nonumber \\& =& 0.
\label{vqnm3}
\end{eqnarray}

\subsection{Near zone analysis}
This equation cannot be solved exactly for the entire range of $1 \leq R < \infty$ due
to the fact that we need to write $r$ as a function of $R$ in (\ref{vqnm3}). Let us now
focus on the near zone, defined by $(r-b) \ll b$. In this region,
\begin{eqnarray}
\ln R &=& (D-3) \ln \lb\frac{r}{b}\rb = (D-3) \ln \lb 1 + \frac{r-b}{b}\rb \nonumber \\
&\sim & (D-3) \lb\frac{r-b}{b}\rb.
\label{vqnm4}
\end{eqnarray}
Thus, in the entire near region, inverting this relation, we have
\begin{equation}
r= b\lb 1 + \frac{\ln R}{D-3} \rb. 
\label{vqnm4a}
\end{equation}

We can therefore approximate $r$ by $b$ in (\ref{vqnm3}) as the 
error $ (r-b) \ll b$. We note that this approximation does not require us to take the large $D$ 
limit explicitly and is valid in the entire near region. This region is expressed in the $R$ coordinate
as $R \ll e^{D-3}$. We will 
not explicitly expand the
function $\Psi_V $ as a series in $1/D$ in the perturbation equation as in \cite{emparan}. Instead, in this section,
we will derive the approximate equation valid in the entire near region. 
Let us now write $\Psi_{V} = R^{\alpha} (R-1)^{\beta} \chi $ where 
\begin{equation}
\alpha = \frac{3}{2}\lb 1 + \frac{1}{D-3} \rb;~~ \beta = \frac{i \omega b}{D-3}.
\label{vqnm5}
\end{equation}
The equation for $\chi$ reduces in the near region to a hypergeometric differential equation, using which we can write in this region;
\begin{eqnarray}
\Psi_V &=& R^{\alpha} (R-1)^{\beta} \left [ c_1 F(a,\tilde b, a+ \tilde b-c+1; 1-R ) \right. \nonumber \\
&+& \left. c_2 (1-R)^{c-a- \tilde b} 
F(c-a, c- \tilde b, c-a- \tilde b +1; 1-R) \right ]; \nonumber \\
a&=& \alpha + \beta - \frac{1}{2(D-3)} + \sqrt{\bar \omega_\ell^{2} - \frac{\omega^2 b^2}{(D-3)^2}}; \nonumber \\
\tilde b &=& \alpha + \beta - \frac{1}{2(D-3)} - \sqrt{\bar \omega_\ell^{2} - \frac{\omega^2 b^2}{(D-3)^2}}; \nonumber \\ 
c &=& 2 \alpha - \frac{1}{D-3}.
\label{vqnm6}
\end{eqnarray}
Here, $\bar \omega_\ell = (\frac{1}{2} + \frac{\ell}{D-3} )$.

We now aim to compute quasinormal modes. As has been observed \cite{emparan}, there are two kinds
of distinct quasinormal modes in the large $D$ limit. A set of modes termed non-decoupled modes are such that the mode frequency
$\omega \sim \mathcal{O}(D)$. 
There is another distinct set of modes that can appear due to
the confining shape of the potential $V_{V}$ in the large
$D$ limit. These modes are ingoing at the horizon and decay far from the horizon and are termed decoupled modes. They have
frequency $\omega \sim \mathcal{O}(1)$ and they
are significant for $b < r < b + \frac{b}{D-3}$, the region where the $t-r$ part of the metric can be 
approximately written in Rindler form. Their significance lies in processes
that are localized in this region very close to the horizon.

Let us first consider the problem of
computing the non-decoupled modes. As has been emphasized many
times in the literature, analytically finding modes that are purely outgoing at infinity or purely
ingoing at the horizon is fraught with difficulties. Quasinormal modes have complex frequencies with a sign of the imaginary part such that they decay in time. As a function of the radial 
coordinate, the solution to the perturbation equation (\ref{vqnm1}) which is outgoing
at infinity is non-normalizable, whereas the ingoing solution at infinity is
exponentially decaying as $r \to \infty$. Further, we have an irregular 
singularity at infinity, and therefore, we only have asymptotic expansions for the 
outgoing and ingoing solutions. Any contamination of the outgoing solution by an 
ingoing part will not change the asymptotic expansions, which are typically in
powers of $(\frac{1}{r})$ whereas the ingoing piece decays exponentially in $r$ 
and therefore faster than any power of $r$. However, such a contamination  
leads to significant differences to the solution when continued to the overlap region in the 
large $D$ approach.
This can affect the values of the quasinormal frequencies. Leaver \cite{leaver}, Sun and Price \cite{sunprice} and 
Nollert and Schmidt \cite{nollert} characterized quasinormal modes as poles of the Laplace transform of the
Green's function for the problem - their procedure eliminates many of the 
computational difficulties. Given the quasinormal mode frequency $\omega$,
formally replace $\omega = - i \Omega$ in the equation ({\ref{vqnm1}). Denote by $f_{+} (\Omega, r_{*})$ the unique solution to (\ref{vqnm1}) 
that is bounded at infinity, and of the form $e^{-\Omega r_{*}}$ as $r_{*} \to
\infty $ for positive
 $\Omega$. Similarly let $f_{-}$ denote the bounded solution 
at the horizon that is of the form
$e^{\Omega r_{*}}$ as $r_{*} \to
- \infty$. It is easy to see that the inverse transformation from $\Omega$ to
$\omega$ takes these boundary conditions to the quasinormal mode ones.
The investigation of the time evolution of an initially bounded 
perturbation of the black hole spacetime is done by taking its Laplace 
transform, and then evaluating the transform using the Green's function.
The (Laplace transform of the ) Green's function can be obtained in a standard
way from $f_{+}$ and $f_{-}$; 
\begin{eqnarray}
G(\Omega, x , x') &=& \frac{1}{W(\Omega )} f_{-}(\Omega, x') f_{+}(\Omega, x), (x' < x); \nonumber \\
&=& \frac{1}{W(\Omega )} f_{-}(\Omega, x) f_{+}(\Omega, x'), (x' > x).
\label{vqnm7}
\end{eqnarray}

This Green's function has poles in the complex $\Omega$
plane at the zeroes of the Wronskian of $f_{-}$ and $f_{+}$, i.e., for those
complex $\Omega$ for which $f_{-}$ and $f_{+}$ are linearly dependent. The 
procedure of taking an inverse Laplace transform by integrating over $\Omega$ to
get the perturbation as a function of time involves an integral over
an appropriate contour in the 
complex $\Omega$ plane. The major contribution to the integral comes from
the poles of the Green's function which are precisely the 
quasinormal modes, and possible branch cuts which contribute to late-time power
law behaviour. This method makes the significance of quasinormal modes clear
in the time evolution of bounded perturbations. As we will show in our problem, it also makes evaluation
of these modes free from errors, as one is computing the zeroes of the 
Wronskian of two functions bounded at either end (horizon or infinity). 

Replacing $\omega = - i \Omega$ in the solution ({\ref{vqnm6}) which is
valid in the near region $r-b \ll b$, the bounded solution at the horizon 
has $c_2 = 0$. We designate this $f_{-}$. The strategy is to work in the 
large $D$ limit, evaluate $f_{+}$ in the far region $r-b >> \frac{b}{D-3}$
and evaluate the Wronskian of the two solutions in the overlap
region $\frac{b}{D-3} << r-b << b$. To this end, we evaluate $f_{-}$ in 
the overlap region. This is obtained from (\ref{vqnm6}) by setting
$c_2  = 0$ and then taking $R$ large.  
We get
\begin{eqnarray}
f_{-} \sim R^{\alpha + \beta} \left[ 
\frac{\Gamma (a+ \tilde b+1-c) \Gamma(\tilde b -a)}{\Gamma(\tilde b) \Gamma(\tilde b -c+1)} 
R^{-a} F\left(a, a-c+1, a- \tilde b +1; \frac{1}{R}\right) \right. \nonumber \\
\left. + \frac{\Gamma (a+\tilde b +1-c) \Gamma(a- \tilde b)}{\Gamma(a) \Gamma(a-c+1)} 
R^{- \tilde b} F\left(\tilde b, \tilde b -c+1, \tilde b -a+1; \frac{1}{R}\right)\right] .
\label{vqnm8}
\end{eqnarray}  

Simplifying, we get the leading large $R$ behaviour
\begin{eqnarray}
f_{-} \sim  \left[
\frac{\Gamma (a+\tilde b +1-c) \Gamma(\tilde b -a)}{\Gamma(\tilde b) \Gamma(\tilde b -c+1)}
R^{\frac{1}{2(D-3)} - \sqrt{\bar \omega_\ell^{2} + \frac{\Omega^2 b^2}{(D-3)^2}}} 
\right. \nonumber \\
\left. + \frac{\Gamma (a+ \tilde b +1-c) \Gamma(a- \tilde b )}{\Gamma(a) \Gamma(a-c+1)}
R^{\frac{1}{2(D-3)} + \sqrt{\bar \omega_\ell^{2} + \frac{\Omega^2 b^2}{(D-3)^2}} } \right] .
\label{vqnm9}
\end{eqnarray}

\subsection{Far region analysis} 
The far region is defined as the region for which $r-b \gg b/(D-3)$. In this limit, the ratio 
$(b/r)^{D-3} \sim C e^{- (D-3)\ln r}$ is a small quantity for both large $D$ and large $r$. Hence we can approximate $f(r) \approx 1$ and $dr_*=dr$.  The far region equation becomes,
\begin{equation}\label{vecfar}
-\frac{d^2\Psi_V}{dr^2}+\frac{(D-3)^2}{4r^2}\left[\left(1+\frac{2\ell}{D-3}\right)^2-\frac{1}{(D-3)^2}\right]\Psi_V=-\Omega^2\Psi_V
\end{equation}
Solutions of this equation are the modified Bessel functions of order $\nu= \frac{D-3}{2}\lb 1+\frac{2\ell}{(D-3)}\rb$.
\begin{equation}\label{vecfarsoln}
\Psi_V=(\Omega r)^{1/2}\ltb d_1 I_\nu(\Omega r)+ d_2 K_\nu(\Omega r)\rtb
\end{equation}

For large $D$ it is more convenient to use a new coordinate $z=\frac{\Omega}{\nu}r$. In terms of $z$, we can use the uniform asymptotic expansion of modified Bessel functions for large order and argument. The asymptotic behaviour of the solutions is $I_\nu(\nu z) \sim e^{\nu z}$ and $K_\nu(\nu z) \sim e^{-\nu z}$. Demanding the solution to be bounded at infinity, we set the coefficient of the growing solution $I_\nu (\nu z)$ to zero.  The expansion of $K_\nu (\nu z)$ is given by

\begin{equation}\label{Kfar}
K_\nu(\nu z)=
\sqrt{\frac{\pi}{2\nu}}\frac{e^{-\nu\eta}}{(1+z^2)^{1/4}}\left[1+\sum_{m=1}^{\infty}(-1)^m\frac{U_m(\tilde{t})}{\nu^m}\right]
\end{equation}

where
\begin{eqnarray}
\eta &=& \sqrt{1+z^2}+\ln\left[\frac{z}{1+\sqrt{1+z^2}}\right]; \nonumber \\
\tilde t &=& \frac{1}{\sqrt{1+z^2}}.
\end{eqnarray}

and $U_m(\tilde t)$ are polynomials in $\tilde t$.
This solution can be extended to the overlap region of the 
near and far regions 
to get the leading order solution for large $D$. The large $D$ 
limit leads, as we recall, to this distinct overlap region 
$\frac{b}{D-3} \ll r-b \ll b$. To find the solution in the overlap region, 
we write $r$ in terms of $R$ by using (\ref{vqnm4a}) which is valid in the
entire near region, and therefore, in particular, in the overlap region.   

The leading order solution in the large $D$ limit in 
the overlap region, denoted by $f_{+}$ is (see Appendix A for details):

\begin{equation}\label{vecfarovlp}
f_{+} = \tilde C R^{\frac{1}{2(D-3)}-\sqrt{\bar \omega_\ell^{2} +\frac{\Omega^2 b^2}{(D-3)^2}}}
\end{equation}

We note that $f_{+}$ is the solution that decays as $r \to \infty$ and it is impossible 
for its asymptotic expansion to be contaminated by the other (linearly independent)
growing solution. 
We also observe that the exponent contains a term 
of order $1/D$, namely $\frac{1}{2(D-3)}$. This also matches exactly with
one of the linearly independent solutions coming from the near region
equation.

If we had worked with the perturbation equation (\ref{vecfar}) with the quasinormal mode frequency $\omega$ and picked the outgoing solution at $\infty$
instead of the Laplace transform approach, our solutions would be given in terms of Bessel functions instead
of modified Bessel functions. The outgoing solution at $\infty$ would be a 
Hankel function.  
For complex frequencies, the outgoing mode is non-normalizable, whereas the
ingoing mode is exponentially decaying. A small contamination of the outgoing mode by an ingoing piece is
unlikely to affect the leading terms in the asymptotic expansion. Indeed, the asymptotic expansion of the 
Hankel function differs in different domains of the complex plane, and in the overlap region of two domains,
we have two different expansions for the Hankel function which differ by exponentially decaying pieces
(c.f p.238-240, \cite{olver}).
However, we have to be careful when continuing the asymptotic expansion to the overlap region of the near
zone and the far region. The ingoing piece is significant in the overlap region. For this reason, 
the Laplace transform approach is preferable.

\subsection{Non-decoupled quasinormal modes}

We have now obtained the form of both the solutions $f_{-}$, bounded at the horizon, and $f_{+}$, 
bounded at $\infty$ in the overlap region, given by (\ref{vqnm9}) and (\ref{vecfarovlp})
respectively. We look for complex values of $\Omega$ when their Wronskian is zero,
i.e., they are linearly dependent in the overlap region. By inspection, it is clear
that one way that this can happen is that the coefficient of the term 
increasing in $R$ in
(\ref{vqnm9}) must go to zero. The coefficient is
$\frac{\Gamma (a+ \tilde b +1-c) \Gamma(a- \tilde b )}{\Gamma(a) \Gamma(a-c+1)}$, which can only go to zero
at the poles of the Gamma functions in the denominator, provided the numerator remains
finite. These cases can be checked, and this possibility ruled out (the Gamma function in the 
numerator also has a pole in this case). The only
other way for $f_{-}$ and $f_{+}$ to be linearly dependent is the limiting case
$\sqrt{\bar \omega_\ell^{2} + \frac{\Omega^2 b^2}{(D-3)^2}} \rightarrow 0$, which
happens for $\Omega b \rightarrow i (D-3) \bar \omega_\ell$. This corresponds 
precisely to the leading order quasinormal mode 
computed in \cite{emparan}, 
$\omega b = (D-3) \bar \omega_\ell $. 
It is important to note that this is true only
in a limiting sense. 
However, it is not possible in this approach to go beyond leading order in $D$ in the computation
of the non-decoupled quasinormal modes. To obtain the corrections to $\Omega$, one would have to deal with coefficient
functions in the near region differential equation of the form $\ln R $. We can put a bound on the $D$ dependence of the correction just by looking at the behaviour of the asymptotic solution pulled to overlap region. 
let $\omega = D \hat \omega$. We can conjecture that next order correction to the quasinormal modes can
be written as $(D-3)[\hat \omega_0 + \hat \omega_{1}/(D-3)^k]$ and look at the behaviour of corrections to the leading order $f_+$ for $\omega_{0} b = (D-3) \bar \omega_\ell $. 
We now want to go beyond leading order in $D$ in the asymptotic expansion for the modified
Bessel function (\ref{Kfar}). 
We see that from the far region, for a range of $k$, the
terms in the asymptotic expansion for the modified Bessel function
 become ill-defined in the overlap region upon plugging
in the leading order value for $\omega_0$ that we have obtained in the large
$D$ limit. Looking at the correction terms,
the necessary condition for the asymptotic
expansion of the modified Bessel to converge is $k \le 2/3 $. (see Appendix A for details).  

Interestingly, the correction beyond leading order has been computed analytically in \cite{emparan}, \cite{emparantanabe} to be $\mathcal{O}(D^{\frac{1}{3}})$,
whereas
numerical results predict a correction $\mathcal{O}(D^{\frac{1}{2}})$ \cite{dias}. The analytical results for the nondecoupled modes, while not done in a $1/D$
expansion implies $k=\frac{2}{3}$ whereas numerical results give $k=\frac{1}{2}$.

\subsection{Decoupled modes}

The decoupled modes $\omega \sim \mathcal{O}(1)$ are ingoing at the horizon and decaying
far from the horizon for large $R$. We take $\ell \sim \mathcal{O}(1)$ as well. From (\ref{vqnm9}), we observe that the solution that is ingoing at the horizon has a piece growing as $R^{\frac{1}{2}}$ in the overlap region
at leading order in $D$. The coefficient of this piece must vanish for the
decoupled quasinormal mode. The two possibilities are 
$a = - m$ or $a-c+1 = -m$ where $m$ is a non-negative integer. There is no
decoupled mode corresponding to the first possibility. But there is 
a mode for $a-c+1 = 0$.
 
We have 
\begin{eqnarray}\label{vqnma}
-\frac{1}{2}-\frac{1}{(D-3)}+\frac{\Omega b}{(D-3)}+\frac{1}{2}\sqrt{\lb 1+\frac{2\ell}{(D-3)}\rb^2+\frac{4\Omega^2b^2}{(D-3)^2}}=-m 
\end{eqnarray}

For $\Omega, l \sim \mathcal{O}(1)$, the square root can be approximated by a series. 

This equation then has a solution only for $m=0$. This is the  decoupled mode 
with frequency upto $\mathcal{O}(\frac{1}{(D-3)})$
\begin{equation}\label{vqnmb}
\omega=i \left [ (\ell-1) + \frac{1}{(D-3)} (\ell-1)^2 \right ] . 
\end{equation}
This answer agrees with past numerical work by Dias, Hartnett and Santos \cite{dias}. It also agrees with the computation of Emparan,Suzuki and Tanabe in 
the $1/D$ expansion \cite{emparan}
to $\mathcal{O}(\frac{1}{(D-3)})$. The next to next to leading order term is similar to that in \cite{emparan} in its dependence on $\ell$ but differs in the numerical value of 
the coefficient. It is $i \frac{2}{(D-3)^2}(\ell-1)^2 $. While we have not used a $1/D$ expansion, our computation uses the approximation $(r-b) << b$ where we neglect terms 
of $\mathcal{O}(\frac{1}{(D-3)^3})$ as they are multiplied by $\ln R $ factors which make the equation hard to solve. The possible impact of these terms in the quasinormal modes is
at $\mathcal{O}(\frac{1}{(D-3)^2})$.
Thus our 
answer yields the quasinormal mode to next-to-leading order.
 
Taking a large $D$ limit of the hypergeometric equation for the mode function, we find
that the parameters in the leading order are such that we have a degenerate case of the hypergeometric 
equation in this limit. The general solution (\ref{vqnm6}) to the complete equation beyond leading order cannot be analytically
expanded about the solution to the leading order degenerate equation (which has a logarithmic singularity) in powers of $1/D$. The problem lies 
with one of the two linearly independent solutions which is {\em outgoing} at the horizon.  
We would like
our solution to be ingoing at the horizon.  
However, upon choosing the ingoing solution, the problem recurs in the 
far limit (\ref{vqnm9}). 
One 
of the Gamma functions in (\ref{vqnm9}) cannot be naively expanded about the leading order
expression in powers of $1/D$. The reason is that the leading order expression is a pole of the Gamma 
function (at leading order, $\tilde b - a = -1$). 
This is a manifestation of the fact that the hypergeometric equation reduces to a degenerate case at
leading order. In \cite{emparan}, in a 
$1/D$ expansion, since the ingoing solution is chosen and
the quasinormal mode identified at the horizon, this does not affect the computation.

\section{Scalar quasinormal modes}

We investigate the equations governing scalar quasinormal modes next. As shown in \cite{ishikodama}, these can be 
reduced to one Schrodinger-type equation
\begin{equation}
\frac{d^2}{dr_{*}^{2}} \Psi_S + (\omega^2 - V_{S} ) \Psi_S = 0.
\label{sqnm1}
\end{equation}

The form of $V_S$ is given in \cite{ishikodama} and it is not possible to solve (\ref{sqnm1}) exactly for all
$r$. We will therefore resort to a near and far region analysis of this equation. We work with the $R$ coordinate instead of $r_{*}$. Then the
equation obeyed by $\Psi_S $ in the near region is of the form
\begin{eqnarray}
& & \frac{d^2}{dR^{2}} \Psi_S + \left [ \frac{1}{R-1} - \frac{1}{(D-3)R} \right ] \frac{d}{dR} \Psi_S \nonumber \\
&+& \left [ \frac{\omega^2 b^2 }{(D-3)^2 (R-1)^2} - \frac{(c_3 R^3 + c_4 R^2 + c_5 R + c_6 )}{4 R^2 (R-1)(pR+q)^2} \right ]\Psi_S
= 0.
\label{sqnm2}
\end{eqnarray}
Here, $c_3, c_4, c_5, c_6, p, q$ are constants that depend on the angular momentum mode $\ell$ and
dimension $D$. 

\begin{align} \label{constants1}
&{p=\frac{2(-1+\ell)}{(D-3)}+\frac{2\left(-1+\ell^2\right)}{(D-3)^2}}\\
&{q=1+\frac{3}{(D-3)}+\frac{2}{(D-3)^2}}
\end{align}
Constants $c_{3---6}$ are given in Appendix B. 

The equation (\ref{sqnm2}) has four regular singular points at $R = -q/p,~0,~1, ~\infty $ and can therefore
be rewritten as a Heun differential equation. However, the solutions to this equation are harder to analyze and in particular, it is difficult to look for solutions satisfying specified boundary conditions at the horizon and infinity.
The reason is that unlike the hypergeometric equation, whose solutions around the singular points are connected by linear transformations, a similar result, termed 
the `connection problem' for the Heun equation is as yet, unsolved except in specific cases.\footnote{For 
some results for the connection problem for adjacent singularities, see \cite{schafkeschmidt}. The singular 
points of interest in our problem, the horizon and $\infty$ are not adjacent; nor can they be made so by
a transformation of R that results in another Heun equation. Hence we cannot use these results.} 

For the decoupled scalar modes, two singular points in the Heun equation merge in the large $D$ limit.
It is hence not possible to compute the quasinormal modes without assuming a 
$1/D$ expansion as has been done in \cite{emparan}. We analyze the equation in detail in Appendix C.

For the nondecoupled modes, instead of working with the Heun equation,
our strategy is to work with a set of three coupled equations which describe the
scalar perturbations. As shown in \cite{ishikodama},these can be reduced to the Heun equation after the use of identities, but we will 
not do so as we would like to circumvent the problem of analyzing a Heun equation. We refer to our past work on
linearized scalar perturbation equations for black strings in (D+1) dimensions (an extra dimension added to the
$D$-dimensional Schwarzschild-Tangherlini metric) which were obtained in \cite{amrutasuneeta}. 
As shown in \cite{amrutasuneeta}, these equations can be simplified to three coupled equations for three 
perturbation variables. We refer the reader to \cite{amrutasuneeta} for a detailed derivation of these equations. Now,
if we consider perturbations independent of the extra dimension along the black string, we recover the 
scalar perturbations of the Schwarzschild-
Tangherlini black hole, and our equations can be used to analyze the non-decoupled scalar quasinormal modes. 
One of the equations decouples in the near region limit, and since its solutions can be written in terms
of hypergeometric functions, it is easy to take the solution to the overlap region (as we did in the case of 
vector perturbations). The coupled equations are for perturbation variables $\hat \psi(r,t), \hat \phi(r,t), 
\hat \eta(r,t)$
which are defined in \cite{amrutasuneeta} in terms of the metric perturbations.

Writing, for example, $\hat{\psi}(r,t)=\psi(r)e^{i \omega t}$, the  system of coupled equations describing
Schwarzschild-Tangherlini (scalar) perturbations is:

\begin{align}\label{psi}
&-\frac{d^2\psi}{dr^2}+\bigg[\frac{(D-2)^3-2(D-2)^2+8(D-2)-8}{4(D-2)r^2}+\frac{f'^2}{4f^2} \nonumber\\
&-\frac{((D-2)^2+2(D-2)-4)}{2(D-2)}\frac{f'}{fr}-\frac{f''}{2f}-\frac{2(D-3)}{(D-2)r^2f}+\frac{k^2}{fr^2}-\frac{\omega^2}{f^2}\bigg]\psi = \nonumber\\
&\left[\frac{4}{f}-\frac{2f'r}{f^2}\right](i\omega)  \eta+\left[\frac{2(D-3)}{(D-3)f}+\frac{2}{(D-2)}-\frac{D}{(D-2)}\frac{rf'}{f}-\frac{r^2 f''}{f}+\frac{f'^2r^2}{2f^2}\right]\phi
\end{align}
\begin{align}
\label{phi}&-\frac{d^2\phi}{dr^2}+\bigg[\frac{(D-2)^3-2(D-2)^2+8(D-2)-8}{4(D-2)r^2}+\frac{f'^2}{4f^2} \nonumber\\
&-\frac{((D-2)^2+2(D-2)-4)}{2(D-2)}\frac{f'}{fr}-\frac{f''}{2f}-\frac{2(D-3)}{(D-2)r^2f}+\frac{k^2}{fr^2}-\frac{\omega^2}{f^2}\bigg]\phi = \nonumber\\
&\frac{2f'}{f^2r}\eta(i\omega) +\left[\frac{2(D-3)}{(D-2)r^4f}-\frac{2(D-3)}{(D-2)r^4}-\frac{2-(D-2)}{(D-2)r^3}\frac{f'}{f}-\frac{f''}{r^2f}+\frac{f'^2}{2f^2r^2}\right]\psi
\end{align}
\begin{align}
\label{eta}&-\frac{d^2\eta}{dr^2}+\left[\frac{(D-2)^2-2(D-2)}{4r^2}-\frac{Df'}{2rf}+\frac{3f'^2}{4f^2}-\frac{3f''}{2f} +\frac{k^2}{fr^2}-\frac{\omega^2}{f^2}\right]\eta \nonumber \\
&\hspace*{5 cm} = \left[\frac{f'}{f}-\frac{2}{r}\right]
\frac{r(i\omega)}{f}\phi-\frac{f'}{f^2}\frac{(i\omega)}{r}\psi
\end{align}

As in the previous cases, we solve these equations in the near and far region separately. In the near region, we rewrite the equations (\ref{psi})---(\ref{eta}) in terms of the $R$ variable using (\ref{vqnm4a}). As we are interested in the non decoupled modes, we consider
$\Omega, \ell \sim \mathcal{O}(D)$ where $\Omega =i \omega$. For simplicity, let us denote $\hat{\Omega}=\frac{\Omega}{(D-3)}$ and $\hat{k}^2=\frac{\ell(\ell+D-3)}{(D-3)^2}$. Taking the large $D$ limit of the equations, 
we see that these equations only decouple at leading order. We define new variables 
\begin{equation}\label{near-HG}
H=\psi+\phi b^2 \text{ and }  G=\psi-\phi b^2.
\end{equation}

The equation for $H$ decouples but unfortunately the equations for $G$ and $\eta$ remained coupled. Let us look at the $H$ equation first.
\begin{equation}\label{near-H-eqn}
\frac{d^2H}{dR^2}+\frac{1}{R}\frac{dH}{dR}-
\left[\frac{1}{4R^2}-\frac{1}{4R^2(R-1)^2}-\frac{1}{R^2(R-1)}+\frac{\hat{k}^2}{R(R-1)}+\frac{\hat{\Omega}^2b^2}{(R-1)^2}\right]H=0
\end{equation}

The solution to this equation is written in terms of hypergeometric functions. Choosing the bounded solution at the horizon, we find
\begin{equation}\label{near-H-sol}
H= C_1 R(R-1)^{\frac{1}{2}+\hat{\Omega}b}F(p,q,1+2\hat{\Omega}b,1-R)
\end{equation}
where,
\begin{align}\label{near-M-pq}
p=\frac{1}{2}\ltb 3+2\hat{\Omega}b-\sqrt{1+4\hat{k}^2+4\hat{\Omega}^2b^2}\rtb \quad
 q=\frac{1}{2}\ltb 3+2\hat{\Omega}b+\sqrt{1+4\hat{k}^2+4\hat{\Omega}^2b^2}\rtb
\end{align}
 We can extend this solution to the overlap region using the standard formulae for hypergeometric functions and taking $R\rightarrow \infty$.
 \begin{equation}\label{near-H-extn}
H=C_1\left[\frac{\Gamma(p+q-2)\Gamma(q-p)}{\Gamma(q)\Gamma(q-2)}R^{\frac{\sqrt{1+4\hat{\Omega}^2b^2+4\hat{k}^2}}{2}}+\frac{\Gamma(p+q-2)\Gamma(p-q)}{\Gamma(p)\Gamma(p-2)}R^{-\frac{\sqrt{1+4\hat{\Omega}^2b^2+4\hat{k}^2}}{2}}\right]
\end{equation}

The equations for $G$ and $\eta$ in the near region are:
\begin{align}
\label{scalarG}&\frac{d^2G}{dR^2}+\frac{1}{R}\frac{dG}{dR}-\left[\frac{1}{4R^2}+\frac{3}{4R^2(R-1)^2}+\frac{1}{R^2(R-1)}+\frac{\hat{k}^2}{R(R-1)}+\frac{\hat{\Omega}^2b^2}{(R-1)^2}\right]G\nn
&\hspace{9 cm}=\frac{4\hat{\Omega}}{R(R-1)^2}\eta b^2\\
\label{scalarEta}&\frac{d^2\eta}{dR^2}+\frac{1}{R}\frac{d\eta}{dR}-\left[\frac{1}{4R^2}+\frac{3}{4R^2(R-1)^2}+\frac{1}{R^2(R-1)}+\frac{\hat{k}^2}{R(R-1)}+\frac{\hat{\Omega}^2b^2}{(R-1)^2}\right]\eta\nn
&\hspace*{9 cm}=\frac{\hat{\Omega}}{R(R-1)^2}G
\end{align}

We cannot solve this system of equations, but using the arguments in \cite{amrutasuneeta} it can be shown that the solution to $G$ and $\eta$ in the overlap region will be of the form
\begin{equation}\label{near-G-extn}
G=\eta=a_1R^{\frac{\sqrt{1+4\hat{\Omega}^2b^2+4\hat{k}^2}}{2}}+a_2R^{-\frac{\sqrt{1+4\hat{\Omega}^2b^2+4\hat{k}^2}}{2}}
\end{equation}

The constants $a_1$ and $a_2$ are both non-zero.

Now let us look at the far region. In this region we have $r\gg b$. Hence we can neglect terms with $f'(r)$ and $f''(r)$ in equations (\ref{psi})---(\ref{eta}) compared to the other $r^{-2}$ terms in the potential. As in the near region, we assume $\Omega^2, k^2$ to be of order $D^2$ and keep only the leading $D$ terms.  We also write $(D-3) \sim D$ etc. at leading order. Approximating $f(r) \approx 1$, we get the far region equations
\begin{align}\label{far-sifi1}
&-\frac{d^2\psi}{dr^2}+\left[\frac{D^2}{4r^2}+\frac{k^2}{r^2}+\Omega^2\right]\psi=2\phi+4\Omega\eta
\\[0.3 cm]
\label{far-sifi2}&
-\frac{d^2\phi}{dr^2}+\left[\frac{D^2}{4r^2}+\frac{k^2}{r^2}+\Omega^2\right]\phi= 0\\[0.3 cm]
\label{far-sifi3}&
-\frac{d^2\eta}{dr^2}+\left[\frac{D^2}{4r^2}+\frac{k^2}{r^2}+\Omega^2\right]\eta=
-2\Omega \phi
\end{align}

The equation for $\phi$ now decouples. The solution for $\phi$ is given in terms of modified Bessel functions. For $\nu = \sqrt{\frac{D^2+1}{4}+k^2}$,

\begin{equation}\label{far-phi-gsol}
\phi=D_1 \sqrt{r}I_\nu(\Omega r)+D_2 \sqrt{r}K_\nu(\Omega r).
\end{equation}

Boundedness as $r \to \infty$  dictates $D_1=0$.  Extending this solution to the overlap region by changing $r$ to $R$, we get
\begin{equation}\label{farphi}
\phi= D_0 R^{-\frac{\sqrt{1+4\hat{k}^2+4\hat{\Omega}^2b^2}}{2}}
\end{equation}

Using this solution as source term, we can find the solution to  $\psi$  and $\eta$ from equations (\ref{far-sifi1}) and (\ref{far-sifi3}). The leading order solution for both turns out to be the same as $\phi$. Taking the combination $(\psi+\phi b^2)$, we calculate zeros of its Wronskian with (\ref{near-H-extn}). The
Wronskian only becomes zero in a limiting sense as 

\begin{equation}
\omega b \rightarrow D\lb \frac{1}{2}+\frac{\ell}{D}\rb
\end{equation} 

We will obtain the same condition by solving the equations for $G$ and $\eta$.
 
For decoupled modes, these equations are difficult to work with since the quasinormal mode is sub-leading in
$D$ and at that order, all the equations (\ref{psi})---(\ref{eta}) are coupled.

\section{Summary and Discussion}

In this paper, we have revisited the problem of linear perturbations of $D$-dimensional Schwarzschild-Tangherlini black holes in the large $D$ limit. 
Following \cite{emparan}, we can define a near zone and a far region with an overlap region which is significant for large $D$. The perturbation equations cannot be solved exactly. They are therefore analyzed in the 
near zone and far region, and the solutions matched in the overlap region in the large $D$ limit. The interest
is in obtaining the quasinormal modes in the large $D$ limit. In previous work, an expansion of the
quasinormal mode frequency and mode function in powers of $1/D$ is used to obtain $\omega \sim \mathcal{O}(1)$. 

We have addressed this problem without the  
expansion of the mode function as a series in $1/D$. Our
results and observations are:
\\
(i) For the vector quasinormal modes $\omega \sim \mathcal{O}(1)$ (decoupled modes), the perturbation equation reduces to a degenerate
case of the hypergeometric equation at leading order in $D$ and its general solution cannot be obtained from that of the
leading order equation as a series in $1/D$. 
We compute these modes without an assumption of a series expansion. We obtain an equation for the decoupled vector quasinormal modes
from which we evaluate the modes to $\mathcal{O}(\frac{1}{(D-3)^2})$. At leading order and the next, it agrees with previous
computations in the $1/D$ expansion \cite{emparan} as well as numerical results \cite{dias}. At next-to-next to leading order, it
agrees in the functional dependence on $\ell$ with \cite{emparan} but differs in the value of the numerical coefficient. Our
computation, while not in a $1/D$ expansion, uses the approximation $(r-b) << b$ and corrections to this possibly affect
results at next-to-next-to leading order. 
\\
(ii) We compute the vector quasinormal modes $\omega \sim \mathcal{O}(D)$ (non-decoupled modes) using a Laplace transform method due to Leaver and others.
We discuss the advantage of this approach while doing matched asymptotic expansions in large $D$. 
However, there are computational issues in using this approach to obtain quasinormal modes beyond leading order.

(iii)We study the equation governing scalar quasinormal modes. This is a Heun equation and hard to
analyze analytically as relatively little is known about solutions to Heun equations in comparison to say, the 
hypergeometric equation.   
For decoupled modes with $\omega \sim \mathcal{O}(1)$,  
an added complication in this problem is that in the large $D$ limit, two of the singular points of the associated Heun equation
are `nearby' and merge in the $D \to \infty$ limit. Due to this reason, it is not possible to compute the decoupled scalar quasinormal
modes without a $1/D$ expansion as done in \cite{emparan}.
For nondecoupled scalar modes, an analysis of the Heun equation is difficult. We instead use an equivalent
set of three coupled perturbation equations we had derived in previous work in the context of a study
of black brane stability \cite{amrutasuneeta}. The equations partly decouple in the near and far regions in the 
large $D$ limit, allowing us to discuss the leading order nondecoupled scalar quasinormal modes in  
this limit.
\\

\section{Appendix A: Expansion of Modified Bessel Functions}

In this section, we describe the procedure to extend the far region solution $f_+=(\Omega r)^{1/2} K_\nu(\Omega r)$ to 
the overlap region in terms of $R$. Here $\nu= \frac{(D-3)}{2}\lb 1+\frac{2\ell}{(D-3)}\rb$. In the case $\Omega \sim \mathcal{O}((D)$, both the amplitude and the order of the modified Bessel function $K_\nu(\Omega r)$ is large. Hence $f_+$ is described using the uniform asymptotic 
expansions for modified Bessel functions. To keep the notation simple, we define a new coordinate $z=\Omega r/\nu=\beta r$.

The large order and large argument expansion of the modified Bessel function is,
\begin{equation}\label{asyexpK}
K_\nu(\nu z)=
\sqrt{\frac{\pi}{2\nu}}\frac{e^{-\nu\eta}}{(1+z^2)^{1/4}}\left[1+\sum_{m=1}^{\infty}(-1)^m\frac{U_m(\tilde{t})}{\nu^m}\right]
\end{equation}

where
\begin{align}
&\eta=\sqrt{1+z^2}+\ln\left[\frac{z}{1+\sqrt{1+z^2}}\right]
&\tilde t=\frac{1}{\sqrt{1+z^2}}
\end{align}
and $U_m(\tilde t)$ are polynomials in $\tilde t$ of degree $3m$.The first three polynomials are 
\begin{align}\label{u_ms}
U_0(\tilde{t})=1, \quad U_1(\tilde{t})=\frac{1}{24}(3\tilde{t}-5\tilde{t}^3), \quad U_2(\tilde{t})=\frac{1}{1152}(81\tilde{t}^2-462\tilde{t}^4+385\tilde{t}^6)
\end{align}

In the large $D$ limit, we can truncate the asymptotic expansion in (\ref{asyexpK}). We retain terms to order $\mathcal{O}(1/D)$ and substitute the
expression for $\eta$ to get

\begin{equation}\label{z_exp}
K_\nu(\nu z)\approx\sqrt{\frac{\pi}{2\nu}}\frac{\left(1+\sqrt{1+z^2}\right)^\nu e^{-\nu\sqrt{1+z^2}}}{z^\nu(1+z^2)^{1/4}}\ltb 1 - \frac{1}{24\nu}\lb\frac{3}{(1+z^2)^{\frac{1}{2}}}-\frac{5}{(1+z^2)^{\frac{3}{2}}}\rb\rtb
\end{equation}

In order to express $K_\nu(\nu z)$ in terms of $R$, we recall $r=bR^{\frac{1}{(D-3)}}$ and in the overlap region, this
can be approximated as
\begin{equation*}
r=b \left[1+\frac{\ln R}{(D-3)}+\frac{(\ln R)^2}{2 (D-3)^2}\right].
\end{equation*}
Note that this expansion is valid only for $\frac{\ln R}{(D-3)} \ll \frac{1}{2}$ which is true in the overlap region.
 
We will now look at each term in (\ref{z_exp}) individually. The term $z^\nu$ is directly proportional to $r^{(D-3)}$. Using  $r^{(D-3)}=b^{(D-3)} R$ we get
\begin{equation}\label{znu in far}
\frac{1}{z^\nu}= \frac{1}{(\beta r)^\nu}= (\beta b)^{-\nu}R^{-\frac{\nu}{(D-3)}}.
\end{equation}
Similarly the prefactor $(\Omega r)^{\frac{1}{2}}$ in $f_+$ is written as $\sqrt{\nu z}=(\nu b)^{\frac{1}{2}} R^{\frac{1}{2(D-3)}}$.
We are interested in finding $f_+$ to the leading order in $D$. Hence we keep terms upto $\frac{\ln R}{(D-3)}$ in expansion of $r$.  The next term in (\ref{z_exp}) becomes

\begin{align}\label{other z nu term}
\left[1+\sqrt{1+z^2}\right]^\nu=&\exp\left[\nu\ln\left\lbrace 1+ \ltb 1 + \beta^2 b^2\lb1+2\frac{\ln R}{(D-3)}\rb\rtb^{1/2}\right\rbrace\right]\nn
=& \lb1+\sqrt{1+\beta^2 b^2}\rb^\nu \exp\ltb \frac{\nu\beta^2 b^2}{ \sqrt{1+\beta^2 b^2} \lb 1+\sqrt{1+\beta^2 b^2}\rb} \frac{\ln R}{(D-3)}\rtb .
\end{align}

Similarly substituting for $z$ we get
\begin{equation}\label{exponential term}
\exp\ltb -\nu\sqrt{1+z^2}\rtb = \lb 1+\beta^2 b^2\rb^{-\nu/2}\exp\ltb \frac{-\nu\beta^2 b^2}{\sqrt{1+\beta^2 b^2}}\frac{\ln R}{(D-3)}\rtb
\end{equation}

The remaining terms in (\ref{z_exp}) can be written in terms of various powers of $(1+z^2)^{-1}$. In the 
overlap region, this can be written as  
\begin{align}\label{series term}
\frac{1}{(1+z^2)}= \lb 1+\beta^2 b^2 \rb^{-1} \ltb1+\frac{\beta^2 b^2}{1+\beta^2 b^2}\frac{2\ln R}{(D-3)}\rtb.
\end{align}

Comparing the coefficients of all the $\ln R$ terms in the above expressions, we see that the contribution from the term $(1+z^2)^{-1/4}$ and the series in (\ref{z_exp}) is smaller by an order $D$ than the other terms. 
For the leading order solution in $D$, we neglect these subleading terms. Substituting all the expressions in $R$ back in (\ref{z_exp}), we get the following expression for $f_+$ (we have absorbed all the constants in one constant $D_0$) :

\begin{equation}\label{vecfar-app}
f_+= D_0 R^{\frac{1}{2(D-3)}-\frac{\nu \sqrt{1+\beta^2 b^2}}{(D-3)}}= D_0 R^{\frac{1}{2(D-3)}-\frac{1}{2}\sqrt{\lb 1+\frac{2\ell}{(D-3)}\rb^2+\frac{4\Omega^2b^2}{(D-3)^2}}}
\end{equation}

In order to calculate the quasinormal modes, we compute the zeroes of the Wronskian of $f_+$ with $f_-$ (obtained from the near region computation). Recall
\begin{equation}\label{vecnear-app}
f_-=c_1R^{\frac{1}{2(D-3)}-\frac{\nu \sqrt{1+\beta^2 b^2}}{(D-3)}}+c_2R^{\frac{1}{2(D-3)}+\frac{\nu \sqrt{1+\beta^2 b^2}}{(D-3)}}
\end{equation}
The constants $c_1$ and $c_2$ depend on $\Omega$ and $\ell$. 

The quasinormal modes obtained by calculating the Wronskian are,
\begin{equation}\label{omega-dir-app}
\lb1+\frac{\Omega^2 b^2}{\nu^2} \rb=0.
\end{equation}
  
After substituting $\nu$, we see that this equation gives us the leading order non-decoupled mode frequency.
We cannot find the value of $\omega$ by calculating the Wronskian beyond leading order due 
to the fact that in the near region, it is computationally difficult to go beyond leading order. But if we assume a next order correction to $\Omega$ as $\Omega=\Omega_0+\frac{\Omega_1}{(D-3)^\alpha}$, we find that the
asymptotic expansion for $f_{+}$ diverges in the overlap region in the large $D$ limit for $\alpha > \frac{2}{3}$. 
To see this, we put back $\beta =\frac{\Omega}{\nu}$ while writing $z$ in terms of $R$. 

\begin{align}\label{series-direct-app}
(1+z^2)&=\ltb 1+\frac{\Omega_0^2b^2}{\nu^2}+\frac{2\Omega_1\Omega_0 b^2}{\nu^2 (D-3)^\alpha}+ \frac{\Omega_0^2b^2}{\nu^2}\frac{2\ln R}{(D-3)}\rtb\nn
&= \lb 1+\frac{\Omega_0^2 b^2}{\nu^2} \rb \ltb 1+\frac{2\Omega_1\Omega_0 b^2}{\nu^2\lb1+\frac{\Omega_0^2 b^2}{\nu^2}\rb(D-3)^\alpha}+ \frac{2\frac{\Omega_0^2 b^2}{\nu^2}}{1+\frac{\Omega_0^2 b^2}{\nu^2}}\frac{\ln R}{(D-3)}\rtb.
\end{align}

This results in the replacement of $\Omega$ in the expression (\ref{omega-dir-app}) by $\Omega_0$. 

To see the effect of leading order matching on the subleading terms of (\ref{z_exp}), let us express the polynomials $U_{k}$ in the coordinate $R$. $U_{k}$ are polynomials in the variable $(\sqrt{1+z^2})^{-1}$. Using (\ref{series-direct-app}) we get (at leading order in the large $D$ limit)
\begin{equation}\label{series-term-app}
\frac{1}{\sqrt{1+z^2}}=\lb 1+\frac{\Omega_0^2 b^2}{\nu^2} \rb^{-\frac{1}{2}} \ltb 1-\frac{\Omega_1\Omega_0 b^2}{\nu^2\lb1+\frac{\Omega_0^2 b^2}{\nu^2}\rb(D-3)^\alpha}- \frac{\frac{\Omega_0^2 b^2}{\nu^2}}{1+\frac{\Omega_0^2 b^2}{\nu^2}}\frac{\ln R}{(D-3)}\rtb.
\end{equation} 

For the $\Omega_0$ obtained in (\ref{omega-dir-app}),$ 1+\frac{\Omega_0^2 b^2}{\nu^2} = 0$. We therefore have to evaluate (\ref{series-term-app}) carefully beyond leading order.

\begin{align}\label{series-term-corrected}
\frac{1}{\sqrt{1+z^2}}&=\ltb \frac{2\Omega_1\Omega_0 b^2}{\nu^2 (D-3)^\alpha}+ \frac{\Omega_0^2b^2}{\nu^2}\frac{2\ln R}{(D-3)}\rtb^{-\frac{1}{2}}\nn
&= (D-3)^{\frac{\alpha}{2}}\lb\frac{2\hat{\Omega}_1\hat{\Omega}_0b^2}{(1+2\hat{\ell})^2}\rb^{-\frac{1}{2}}\ltb 1+\frac{\hat{\Omega}_0}{\hat{\Omega}_1}\frac{\ln R}{(D-3)^{1-\alpha}}\rtb^{-\frac{1}{2}}
\end{align} 

The hatted quantities denote $\hat{\Omega}=\frac{\Omega}{(D-3)}$ and $\hat{\ell}=\frac{\ell}{(D-3)}$.
As we are in the overlap region, $1\ll \ln R \ll \frac{(D-3)}{2}$.  Let us now look at the subleading term in (\ref{z_exp}) .
The  series term $U_1$ after substituting \ref{series-term-corrected} is,
\begin{align}\label{U1-NLO-app}
&\frac{1}{\nu}\bigg\lbrace U_1\lb\frac{1}{\sqrt{1+z^2}}\rb\bigg\rbrace =\nn
 &\frac{(D-3)^{\frac{\alpha}{2}}(1+2\hat{\ell})}{(D-3)}\lb\frac{2\hat{\Omega}_1\hat{\Omega}_0b^2}{(1+2\hat{\ell})^2}\rb^{-\frac{1}{2}}\frac{1}{8}\ltb 1+\frac{\hat{\Omega}_0}{\hat{\Omega_1}}\frac{\ln R}{(D-3)^{1-\alpha}}\rtb^{-\frac{1}{2}}-\nn&\frac{(D-3)^{\frac{3\alpha}{2}}(1+2\hat{\ell})}{(D-3)}\frac{5}{24}\lb\frac{2\hat{\Omega}_1\hat{\Omega}_0b^2}{(1+2\hat{\ell})^2}\rb^{-\frac{3}{2}}\ltb1+\frac{\hat{\Omega}_0}{\hat{\Omega_1}}\frac{\ln R}{(D-3)^{1-\alpha}}\rtb^{-\frac{3}{2}}
\end{align}

Consider the $D$ dependence of the second term.  The coefficient of the second  term on (\ref{U1-NLO-app}) is proportional to $(D-3)^{\frac{3\alpha-2}{2}}$. For $\alpha >\frac{2}{3}$, the exponent $\frac{3\alpha-2}{2}$ becomes positive.
We can further see that from the expressions for (\ref{u_ms}), for each $m$ the series terms in (\ref{asyexpK}) ,
the leading term will be $\approx (D-3)^{\frac{(3 \alpha-2)m}{2}}$.
For $\alpha >\frac{2}{3}$, $(3 \alpha-2)m$ is positive and both the series diverge for large $(D-3)$. For $\alpha=\frac{2}{3}$, the leading order term for each $m$ is $\mathcal{O}(1)$. Hence we cannot truncate the
asymptotic expansion for large $D$. 
The asymptotic expansion of the solution $f_{+}$ thus breaks down for $\alpha>\frac{2}{3}$. 
Interestingly, $\alpha =  \frac{2}{3}$ corresponds to the correction $\omega_{1} \sim \mathcal{O}(D^{1/3})$ obtained in \cite{emparan}.
The value $\alpha=\frac{1}{2}$ found numerically in \cite{dias} is also
consistent with convergence of the asymptotic expansion.

\section{Appendix B: Constants in the scalar perturbation equation}
Here, we give the values of various constants in the equation (\ref{sqnm2}). We denote $(D-3)=d$.
\begin{align} \label{constants2}
c_3=&\frac{4 (-1+\ell)^2}{d^2}+\frac{8 \left(1+\ell-5 \ell^2+3 \ell^3\right)}{d^3}+\frac{4 \left(10 \ell-7 \ell^2-16 \ell^3+13 \ell^4\right)}{d^4}+\nonumber \\
&{\frac{8 \left(-1+3 \ell+5 \ell^2-9 \ell^3-4 \ell^4+6 \ell^5\right)}{d^5}+\frac{4 \left(-1+6 \ell^2-9 \ell^4+4 \ell^6\right)}{d^6}}\\
c_4=&-\frac{12 (-1+\ell)}{d}-\frac{12 \left(-3+\ell+2\ell^2\right)}{d^2}-\frac{12 \left(-4+3 \ell-\ell^2+2 \ell^3\right)}{d^3}-\nonumber\\
&{\frac{12 \left(-4+7 \ell-4 \ell^3+\ell^4\right)}{d^4}+\frac{12 \left(3-4 \ell-7 \ell^2+6 \ell^3+2 \ell^4\right)}{d^5}+}\nonumber\\
&{\frac{12 \left(1-4 \ell^2+3 \ell^4\right)}{d^6}}\\
c_5=&1+\frac{2 (-3+4 \ell)}{d}+\frac{4 \left(-11+7 \ell+2 \ell^2\right)}{d^2}+\frac{2 \left(-45+20 \ell+14 \ell^2\right)}{d^3}\nonumber\\
&+\frac{-89+44 \ell+40 \ell^2}{d^4}+\frac{4 \left(-12+6 \ell+11 \ell^2\right)}{d^5}+\frac{12 \left(-1+2 \ell^2\right)}{d^6}\\
c_6=&1+\frac{8}{d}+\frac{26}{d^2}+\frac{44}{d^3}+\frac{41}{d^4}+\frac{20}{d^5}+\frac{4}{d^6}
\end{align}

\section{Appendix C: Scalar decoupled quasinormal modes }
We analyze the equation for the decoupled modes.
To get the $R$ equation in the form of a Heun equation we define
\begin{equation}\label{Psi ansatz}
\Psi_S=R^{\alpha}(R-1)^{\beta}(pR+q)^{\gamma}\chi
\end{equation}
The equation for $\chi$ is
\begin{equation}\label{Heun in R}
\frac{d^2 \chi}{dR^2}+\ltb \frac{2\alpha+d_1}{R}+\frac{1+2\beta}{R-1}+\frac{2\gamma}{R+\frac{q}{p}}\rtb\frac{d \chi}{d R}+\frac{AB R-C}{R(R-1)(R+\frac{q}{p})}\chi=0
\end{equation}
Here,
\begin{align}
\alpha=\frac{1}{2}+\frac{1}{2(D-3)} \qquad
\beta=\frac{i \omega b}{(D-3)} \qquad \gamma = 2 \qquad d_1=-\frac{1}{(D-3)}
\end{align}

The constants $A,B,C$ are
\begin{align}
&A+B = 2(\alpha+\beta+\gamma)+d_1\\
&AB = (\alpha+\gamma)(1+2\beta)+2\alpha\gamma+(\gamma+\beta) d_1+\frac{(c_5+c_6)}{4 p q}-\frac{c_6}{2 q^2}-\frac{(c_3+c_4+c_5+c_6)}{4p(p+q)}\\
&C=-\frac{\alpha(1+2\beta)q}{p}+2\alpha\gamma-\frac{\beta d_1 q}{p}+\gamma d_1+\frac{c_5+c_6}{4pq}-\frac{c_6}{2q^2}
\end{align}

At leading order in $D$, the various constants in the equation can be evaluated, and $p \to 0$ in this limit
($q \to 1$). This results from setting $\ell \sim \mathcal{O}(1)$ for decoupled modes. The singular points in the Heun equation are at  $R = -q/p,~0,~1,~\infty $, and in this limit,
the singular point at $-q/p$ approaches the point at infinity.\footnote{For non-decoupled modes for which $\ell \sim \mathcal{O}(D)$, we do not have $p \to 0$.} We now examine this limit carefully. We will
employ a rescaling that is often used to obtain the confluent Heun equation from the Heun equation.
We define a new coordinate $\bar R$ by $R = \frac{-q}{p} \bar R $. We also define the constant $\bar C$
by $C =  \frac{-q}{p} \bar C$. We first perform the rescaling and then take the limit $p \to 0$ and $q \to 1$ (the values
at leading order in $D$). This yields the equation
\begin{equation}
\label{hyplimit}
\frac{d^2 \chi}{d\bar R^2} + \ltb \frac{2\alpha+d_1 + 1 + 2 \beta}{\bar R} + \frac{2\gamma}{\bar R-1} \rtb \frac{d \chi}{d \bar R} + \frac{AB \bar R- \bar C}{\bar R^2 (\bar R-1)}\chi=0.
\end{equation}

However, we also need to evaluate all the constants in this equation in the large $D$ limit. We find that in this limit,
$\bar C = 0$ in (\ref{hyplimit}). The equation (\ref{Heun in R}),upon taking the limit $D \to \infty$ reduces to a
hypergeometric equation
\begin{equation}
\label{hyplimit1}
\frac{d^2 \chi}{d\bar R^2} + \ltb \frac{2\alpha+d_1 + 1 + 2 \beta}{\bar R} + \frac{2\gamma}{\bar R-1} \rtb \frac{d \chi}{d \bar R} + \frac{AB }{\bar R (\bar R-1)}\chi=0.
\end{equation}

It must be noted that all the constants in (\ref{hyplimit1}) take their leading order in $D$ values, and do not contain
the quasinormal mode frequency $\omega$ which is not of $\mathcal{O}(D)$. Thus this equation cannot be used to obtain
the decoupled mode frequency $\omega$ as this equation is valid only
at leading order.

For any finite $D$, in the Heun equation (\ref{Heun in R}), the singular point at $R= - q/p$ has not actually merged with the singular point at infinity.
Rather, they are `nearby' and merge in the large $D$ limit. To compute decoupled modes with
$\omega \sim \mathcal{O}(1)$, if they are present , we need to take into account sub-leading corrections in $D$ in the various constants
in (\ref{Heun in R}).
We re-write the equation in terms of $z=\frac{1}{R}$. In the new coordinates, the black hole horizon lies at $z=1$ and infinity is mapped to $z=0$.

The singularity at $z=-\frac{p}{q}$ tends to zero in the decoupled mode case in the large $D$ limit. To get the equation (\ref{Heun in R}) in canonical Heun form in terms of $z$, we define $\chi=z^\delta\Phi$. Then, for an appropriate value of
$\delta$, $\Phi$ obeys a Heun equation
\begin{equation}
\label{Heun in z}
\frac{d^2 \Phi}{dz^2}+\ltb\frac{2\delta + 1 -2(\alpha+\beta+\gamma)-d_1}{z}+\frac{1+2\beta}{z-1}+\frac{2\gamma}{z+\frac{p}{q}}\rtb\frac{d \Phi}{dz}+\frac{(MN z-L)\Phi}{z(z-1)\lb z+\frac{p}{q}\rb}=0
\end{equation}
In terms of the constants in the original equation (\ref{Heun in R}),
\begin{equation}
\delta= \frac{1}{2}\ltb 2(\alpha+\beta+\gamma)+d_1+\sqrt{(2(\alpha+\beta+\gamma)+d_1)^2-4AB}\rtb
\end{equation}
\begin{equation}
L=-\frac{p}{q}C+\delta\lb 2\gamma-\frac{p(1+2\beta)}{q}\rb-AB\lb1-\frac{p}{q}\rb
\end{equation}
\begin{equation}
M=\delta; \quad N=\delta +1-(2\alpha + d_1)
\end{equation}

We are interested in finding those quasinormal mode frequencies for which the mode function is ingoing at the
horizon and bounded at $\infty $ (since we are working with the near region equation,
we are interested in solutions that decay for large $R$). We let $\Omega = i \omega$ and map an ingoing solution at
the horizon to a bounded solution. Then, we are interested in finding $\Omega$ for which the solution to
(\ref{Heun in z}) is bounded both at $z=0$ and $z=1$. The complication is that $\Omega$ is sub-leading in $D$, but
in the large $D$ limit, the singular point at $z = -\frac{p}{q} \to 0$ and merges with the singular point at
$z=0$. Unfortunately, results on merging singular points of the Heun equation use an expansion of the solution
as a series of hypergeometric functions. At leading order in $D$, some parameters of these hypergeometric functions
become degenerate and it is not possible to use these results. We are thus unable to evaluate quasinormal
modes as was done for the vector case. The decoupled scalar quasinormal modes have been evaluated in a
$1/D$ expansion in \cite{emparan}. This reproduces past numerical results for these modes in \cite{dias} at finite
(large) $D$ at leading order and computes higher order corrections.

\end{document}